\documentclass[aps,a4paper, preprint]{revtex4-1}
\usepackage{graphicx}
\usepackage{amsmath}
\usepackage{amssymb}
\usepackage{times}
\usepackage{ulem}
\usepackage[utf8]{inputenc}
\usepackage{placeins}
\usepackage{makecell}
\usepackage[version-1-compatibility]{siunitx}
\usepackage[pdfborder=0]{hyperref}
\hypersetup{
    colorlinks=true,
    linkcolor=blue,          
    citecolor=blue           
}

\def\beq{\begin{equation}}
\def\eeq{\end{equation}}
\def\beqna{\begin{eqnarray}}
\def\eeqna{\end{eqnarray}}
\def\bea{\begin{array}}
\def\ea{\end{array}}
\def\mg{{\mathcal G}}
\def\MU{{\mathcal U}}
\def\MV{{\mathcal V}}

\def\Re{\operatorname{Re}}
\def\Im{\operatorname{Im}}
\def\etal{{\it et al.~}}

\begin{document}
\title{Feedback-enhanced squeezing or cooling of fluctuations in a parametric
resonator}
\author{Adriano A. Batista}
\affiliation{
Departamento de Física, Universidade Federal de Campina Grande\\
Av. Apr\'{i}gio Veloso 882,
Campina Grande-PB, CEP: 58.429-900, Brazil
}
\date{\today}
\begin{abstract}
Here we analyse ways to achieve deep subthreshold parametric squeezing of
fluctuations beyond the $-6$~dB limit of single degree-of-freedom parametric
resonators.
One way of accomplishing this is via a lock-in amplifier feedback loop.
Initially, we calculate the phase-dependent parametric amplification
with feedback of an added ac signal.
In one approach, we use the averaging method to obtain the amplification gain,
while in the second approach, we obtain the ac response of the parametric
amplifier with feedback using the harmonic balance method. 
In this latter approach, the feedback is proportional to an integral term that
emulates the cosine quadrature output of a lock-in amplifier multiplied by a
sine at the same tone of the lock-in.
We find that the gain obtained via these two methods are the same whenever the
integration time span of the integral is a multiple of the tone period.
When this is not the case, we can obtain considerable deamplification.
Finally, we analyse the response of the parametric resonator with
feedback, described by this integro-differential model, to an added white noise
in the frequency domain.
Using this model we were able to calculate, in addition to squeezing, the
noise spectral density in this resonator with feedback.
Very strong squeezing or cooling can be obtained.
\end{abstract}
\maketitle
\section{Introduction}
Parametrically driving resonators play an
important role in nanomechanical systems \cite{bachtold2022mesoscopic}.
Driving parametrically a resonator is a way to obtain very high effective
quality factors \cite{miller2018effective}.
Hence, amplifiers based on these resonators can achieve very high gains and a
very narrow gain bandwidth.
Ths narrow band decreases the effect of fluctuations due to added noise.
Very sensitive accelerometers \cite{zhao2019toward}, force \cite{moser2013ultrasensitive} and mass sensors \cite{zhang2005application, papariello2016ultrasensitive} based on these resonators have been
implemented experimentally in micro and nanomechanical systems.

Further decrease of fluctuations can be achieved when  noise squeezing
techniques are used.
In a pioneering paper in 1991, Rugar and Grütter \cite{rugar91} experimentally
observed thermal noise squeezing in a capacitively actuated
 micromechanical parametric resonator. 
They experimentally obtained $-4.9$~dB of squeezing and theoretically predicted
a lower bound of $-6$~dB noise squeezing at the parametric instability
threshold with the parametric pump frequency set at twice the fundamental mode
of the resonator.
Only in 2013 this lower bound was surpassed by a lock-in feedback scheme
developed by Vinante and Falferi in Ref.~\cite{vinante2013feedback}. 
In this scheme, a -11.3~dB in squeezing of thermal
fluctuations of a microscopic silicon beam cantilever was reached experimentally.
Poot \etal \cite{poot2015deep} used the same squeezing technique with
parametric pump and feedback that achieved noise compression at -15.1 dB in
noise power reduction in an electro-optomechanical system.
The same type of feedback scheme was applied to obtain strong squeezing in an
optomechanical membrane \cite{sonar2018strong}, no new theoretical analysis was
developed.
The same feedback scheme and a similar analysis of the data as in Vinante and Falferi paper seems
to have been used in Ref.~\cite{mashaal2024strong}.
Each one of these papers \cite{vinante2013feedback, poot2015deep, sonar2018strong, mashaal2024strong} makes important experimental advancements in the
implementation of squeezing enhanced by feedback, nevertheless, a consistent
stochastic theory of fluctuations squeezing with feedback is still missing.

Here we extend our previous work on squeezing of fluctuations in
parametrically-modulated resonators with added noise \cite{batista2024deep} by
including feedback.
Specifically, we analyse and generalize the linear feedback scheme proposed by
Vinante and Falferi to enhance squeezing of fluctuations.
We believe our approach is more rigorous and more complete since we
take into account detuning, we avoid using the averaging method to
analyse the fluctuations in the frequency domain, but instead we
Fourier transform the stochastic integro-differential equations of our model
into an algebraic stochastic system in the frequency domain.
We were able to obtain the dispersions in the sine and cosine quadratures
at half the pump frequency.
We also find that correlation arises between these quadratures when there is
detuning between half the pump frequency and the natural frequency of the
resonator.
Additionally, we also calculate the noise spectral density (NSD) of the resonators fluctuations.
When the parametric resonator with feedback is excited by an added ac signal,
without noise, we obtain the gain curve as a function of phase
using the harmonic balance method.
This analysis of a deterministic stationary response complements our stochastic
analysis of squeezing.
It is a simple way to check our results on squeezing.

The remainder of this paper is organized as follows.
In Sec.~\ref{theory} we initially analyse parametric
amplification with two feedback schemes, and
later, we propose and analyse an stochastic integro-differential model of
enhanced fluctuations squeezing, cooling, and calculate the noise spectral density (NSD). 
In Sec.~\ref{results} we analyse and discuss our numerical results, and in
Sec.~\ref{conclusion} we draw our conclusions.
\section{Theory}
\label{theory}
We will initially investigate the equation of motion of a
parametrically-modulated resonator with lock-in feedback as proposed in
Ref.~\cite{vinante2013feedback}, but driven by an external ac drive instead of
white noise.
This equation is given by
\beq
\ddot x+\gamma \dot x+\omega_0^2x-F_p\sin(2\omega t)x=\eta u\sin\omega
t+F_s\sin(\omega t+\phi),
\label{polf_gain}
\eeq
where $\gamma$ is the dissipation rate, $\omega_0$ is the angular natural 
frequency of the resonator, $F_p$ is the pump amplitude, $2\omega$ is the pump
frequency, $\eta$ is the feedback constant, and $u$ is the cosine quadrature
of $x(t)$. 

The alternative model of feedback that we propose is given by the following
integro-differential equation
\beq
\ddot x+\gamma \dot x+\omega_0^2x-F_p\sin(2\omega t)x=
\frac{2\eta}\tau \int_{t-\tau}^t x(t')\cos(\omega t')dt'\sin\omega t
+F_s\sin(\omega t+\phi),
\label{polf_gain_integral}
\eeq
where $\tau$ is the integration time span of a lock-in amplifier (it is also
known as the time constant of the lock-in).
We will see below that when $\tau$ is an integer multiple of $2\pi/\omega$ we
recover the same gain obtained in the amplifier described by
Eq.~\eqref{polf_gain}.
\subsection{Phase-dependent amplification in the parametrically-driven resonator with lock-in feedback: averaging approach}
Let us analyse the gain dependence on phase obtained from
Eq.~\eqref{polf_gain} in a similar way to what was performed by Rugar and
Grütter in Ref. \cite{rugar91}.
This analysis gives us insight on the response of this system to added noise.
We now transform the fast variables $(x(t), \dot x(t))$ from
Eq.~\eqref{polf_gain} to the slow variables $(\MU(t), \MV(t))$ via 
$x(t)= \MU(t)\cos\omega t-\MV(t)\sin\omega t$ and
$\dot x(t)= -\omega\left[\MU(t)\sin\omega t+\MV(t)\cos\omega t\right]$.
Subsequently, we apply the averaging method to the equations of motion for
$\MU(t)$ and $\MV(t)$ and obtain the following autonomous dynamical system
\beq
\begin{aligned}
\dot u &=
-\frac\gamma2u-\frac1{2\omega}\left[\left(\frac{F_p}2+\eta\right)u+\Omega v
+F_{s}\cos\phi\right],\\
\dot v &= -\frac\gamma2 v+\frac1{2\omega}\left(\Omega
u+\frac{F_p}2v-F_{s}\sin\phi\right),
\end{aligned}
\label{uv_avg}
\eeq
where $\MU(t)\approx u(t)$ and $\MV(t)\approx v(t)$.
For more details on the application of the averaging method see
Ref.~\cite{batista2012heating}.
We can write the fixed-point solution of Eq.~\eqref{uv_avg} as
\beq
\left(
\bea{c}
u\\v
\ea
\right)
=\frac{F_s}{\left(\frac{F_p}2+\gamma\omega
+\eta\right)\left(\frac{F_p}2-\gamma\omega\right)-\Omega^2}
\left(
\bea{cc}
\frac{F_p}2-\gamma\omega&-\Omega \\
 -\Omega &\frac{F_p}2+\gamma\omega +\eta
 \ea
\right)
\left(
\bea{c}
-\cos\phi\\
\sin\phi
\ea
\right)
\eeq
Consequently, we can write the stationary squared amplitude 
$r^2=u^2+v^2$ as
\beq
\frac{r^2}{F_s^2}=
\frac{\Omega^2+\frac{F_p^2}4+\gamma^2\omega^2+\frac{\eta^2}2+\eta\left(\frac{F_p}2+\gamma\omega\right)-\left[F_p\gamma\omega+\frac\beta2\left(\eta+F_p+2\gamma\omega\right)\right]\cos2\phi+\Omega(F_p+\eta)\sin2\phi}
{\left[\left(\frac{F_p}2+\gamma\omega
+\eta\right)\left(\frac{F_p}2-\gamma\omega\right)-\Omega^2\right]^2}.
\label{deep_squeeze}
\eeq
Based on Eq.~\eqref{deep_squeeze}, we obtain that the instability threshold of parametric oscillator with lock-in feedback is given by
\beq
\frac{F_p^2}4-\gamma^2\omega^2-\Omega^2 =-\eta\left(\frac{F_p}2-\gamma\omega\right).
\label{threshold}
\eeq
From Eq.~\eqref{deep_squeeze}, we find the minimum and maximum amplitude of
the response of the parametric amplifier at zero detuning ($\Omega=0$) to be
\beq
\begin{aligned}
r_{min}^2 &= \frac{F_s^2}{\left(F_p/2+\gamma\omega+\eta\right)^2},\\
r_{max}^2 &= \frac{F_s^2}{\left(F_p/2-\gamma\omega\right)^2}.
\end{aligned}
\label{r_min_r_max}
\eeq

\subsection{Phase-dependent amplification in the parametrically-driven resonator with lock-in feedback: harmonic balance approach}
We now perform an analysis similar to the one developed in the previous
subsection to obtain the phase-dependent gain of the driven parametric
resonator with feedback whose dynamics is described by
Eq.~\eqref{polf_gain_integral}.
We seek a stationary solution to this integro-differential equation in the form
\beq
x(t)=\frac12\left[A(\omega)e^{-i\omega t}+A^*(\omega)e^{i\omega t}\right].
\eeq
Using the fact that the functions $e^{\pm i\omega t}$ are linearly independent,
we find
\beq
\begin{aligned}
&(\omega^2-\omega_0^2-i\gamma\omega)A(\omega)-\frac{iF_p}2A^*(\omega)-\frac{i\eta}2(A+A^*)-\frac{\eta A}{8\tau\omega}\left(1-e^{2i\omega\tau}\right)=iF_se^{-i\phi},\\
&\left[1-\frac{i\eta\chi(\omega)}2-\frac{\eta\chi(\omega)}{4\tau\omega}\left(1-e^{2i\omega\tau}\right)\right]A(\omega)-\frac{i(F_p+\eta)\chi(\omega)}2A^*(\omega)=i\chi(\omega)F_se^{-i\phi},
\end{aligned}
\eeq
which can be rewritten in the more compact form
\beq
\begin{aligned}
&a(\omega)A(\omega)-ib(\omega)A^*(\omega)=i\chi(\omega)F_se^{-i\phi},\\
&ib^*(\omega)A(\omega)+a^*(\omega)A^*(\omega)=-i\chi^*(\omega)F_se^{i\phi},
\end{aligned}
\label{eq:A_om}
\eeq
where
\[
\begin{aligned}
a(\omega)
&=1-\frac{\eta\chi(\omega)}{2\tau}\left[\frac{\left(1-e^{2i\omega\tau}\right)}{2\omega}+i\tau\right],\\
b(\omega) &=\chi(\omega)\frac{F_p+\eta}{2}.
\end{aligned}
\]
Solving the algebraic system \eqref{eq:A_om}, we find
\beq
A(\omega) = \frac{ia^*(\omega)\chi(\omega)e^{-i\phi}+b\chi^*(\omega)e^{i\phi}}{|a(\omega)|^2-|b(\omega)|^2}F_s.
\label{AomF_s}
\eeq
If $\tau$ is a multiple of $2\pi/\omega$ we recover the amplification gain
obtained with the averaging method in Eq.~\eqref{deep_squeeze}.

\subsection{Parametrically-driven resonator with lock-in feedback and added noise}
The Langevin equation of a damped parametrically driven oscillator with
added noise and with feedback from the cosine quadrature output of a lock-in  
is given by
\beq
\ddot x+\gamma \dot x+\omega_0^2x-F_p\sin(2\omega t)x=
\frac{2\eta}\tau \int_{t-\tau}^t x(t')\cos(\omega t')dt'\sin\omega t
+r(t).
\label{polf_noise}
\eeq
By Fourier transforming Eq.~\eqref{polf_noise}, we obtain
\beq
\begin{aligned}
\tilde x(\nu)&=\chi(\nu)\tilde r(\nu)
+\frac{F_p\chi(\nu)}{2i}[\tilde x(\nu+2\omega)-\tilde x(\nu-2\omega)]\\
&+\frac{\eta\chi(\nu)}{\tau}\int_{-\infty}^\infty
\left[\frac{e^{i(\nu+\omega)t}}{\nu+\omega}-\frac{e^{i(\nu-\omega)t}}{\nu-\omega}\right]
\left\{x(t)\cos(\omega t)-x(t-\tau)\cos[\omega(t-\tau)]\right\}dt\\
&=\chi(\nu)\tilde r(\nu)
+\frac{F_p\chi(\nu)}{2i}[\tilde x(\nu+2\omega)-\tilde x(\nu-2\omega)]\\
&+\frac{\eta\chi(\nu)}{2\tau}
\left[\frac{\tilde x(\nu+2\omega)+\tilde x(\nu)}{\nu+\omega}-\frac{\tilde
x(\nu)+\tilde x(\nu-2\omega)}{\nu-\omega}\right]\\
&-\frac{\eta\chi(\nu)}{2\tau}\left\{\frac{e^{i(\nu+\omega)\tau}\left[\tilde x(\nu+2\omega)+\tilde x(\nu)\right]}{\nu+\omega}
-\frac{e^{i(\nu-\omega)\tau}\left[\tilde x(\nu)+\tilde
x(\nu-2\omega)\right]}{\nu-\omega}\right\}\\
&=\chi(\nu)\tilde r(\nu)
+\frac{F_p\chi(\nu)}{2i}[\tilde x(\nu+2\omega)-\tilde x(\nu-2\omega)]\\
&+\frac{\eta\chi(\nu)}{2\tau}\left\{\frac{\left[1-e^{i(\nu+\omega)\tau}\right]\left[\tilde x(\nu+2\omega)+\tilde x(\nu)\right]}{\nu+\omega}
-\frac{\left[1-e^{i(\nu-\omega)\tau}\right]\left[\tilde x(\nu)+\tilde x(\nu-2\omega)\right]}{\nu-\omega}\right\}
,
\label{polf_noise_fourier}
\end{aligned}
\eeq
where we use the shorthand notation
\[
\chi(\omega)=\frac1{\omega_0^2-\omega^2-i\gamma\omega}.
\]
Here, we use the following notation for the Fourier transform
\[
\tilde f(\nu)=\int_{-\infty}^\infty e^{i\nu t} f(t)\,dt.
\]
\subsubsection{Squeezing with feedback}
At $\nu=\omega$, by neglecting off-resonance terms with $\tilde x(\pm3\omega)$
in Eq.~\eqref{polf_noise_fourier}, we obtain 
\beq
\begin{aligned}
\left\{1-\frac{\eta\chi(\omega)}{2\tau}\left[\frac{\left(1-e^{2i\omega\tau}\right)}{2\omega}
+i\tau\right]\right\}\tilde x(\omega)
-i\chi(\omega)\frac{\eta+F_p}{2}\tilde x^*(\omega)
&=\chi(\omega)\tilde r(\omega),\\
i\chi^*(\omega)\frac{\eta+F_p}{2}\tilde x(\omega)
+\left\{1-\frac{\eta\chi^*(\omega)}{2\tau}\left[\frac{\left(1-e^{-2i\omega\tau}\right)}{2\omega}
-i\tau\right]\right\}\tilde x^*(\omega)
&=\chi^*(\omega)\tilde r^*(\omega),
\end{aligned}
\label{noise_response}
\eeq
where we neglected the off-resonance terms with $\tilde x(3\omega)$.
Solving this, we find
\beq
\tilde x(\omega) = \frac{a^*(\omega)\chi(\omega)\tilde
r(\omega)+ib\chi^*(\omega)\tilde r^*(\omega)}{|a(\omega)|^2-|b(\omega)|^2},
\label{xom_fb}
\eeq
where
\[
\begin{aligned}
a(\omega)
&=1-\frac{\eta\chi(\omega)}{2\tau}\left[\frac{\left(1-e^{2i\omega\tau}\right)}{2\omega}+i\tau\right],\\
b(\omega) &=\chi(\omega)\frac{F_p+\eta}{2}.
\end{aligned}
\]
For the special case in which $\tau$ is an integer multiple of $2\pi/\omega$,
this can be simplified to
\beq
\begin{aligned}
a(\omega)
&=1-\frac{i\eta\chi(\omega)}{2},\\
b(\omega) &=\frac{F_p+\eta}{2}\chi(\omega).
\end{aligned}
\eeq
In this feedback scheme, we find that the instability threshold ($|a(\omega)|=|b(\omega)|$) is the same as the one obtained by the averaging method in Eq.~\eqref{threshold}.

The resonator response $\tilde x(\omega)$ given in Eq.~\eqref{xom_fb} can be
rewritten as
\beq
\tilde x(\omega)=\tilde G_0(\omega)\tilde r(\omega)+\Gamma(\omega)\tilde
r^*(\omega),
\label{noise_response_fb}
\eeq
where
\beq
\begin{aligned}
\tilde G_0(\omega)
&=\frac{a^*(\omega)\chi(\omega)}{|a(\omega)|^2-|b(\omega)|^2}
=\frac{\Omega+i(\gamma\omega+\eta/2)}{\Omega^2+\gamma^2\omega^2-F_p^2/4+\eta(\gamma\omega-F_p/2)}\\
\Gamma(\omega) &=\frac{ib(\omega)\chi^*(\omega)}{|a(\omega)|^2-|b(\omega)|^2}
=\dfrac{i(F_p+\eta)/2}{\Omega^2+\gamma^2\omega^2-F_p^2/4+\eta(\gamma\omega-F_p/2)}.
\end{aligned}
\eeq
In the following, we use the approach developed in Ref.~\cite{batista2024deep}
to analyse squeezing of fluctuations.
From Eq.~\eqref{noise_response_fb}, we find
\beq
\begin{aligned}
    \tilde x'(\omega) &=\tilde G_0'(\omega)\tilde r'(\omega)-\tilde G_0''(\omega)\tilde r''(\omega)
+\Gamma'(\omega)\tilde r'(\omega)+\Gamma''(\omega)\tilde r''(\omega),\\
    \tilde x''(\omega) &=\tilde G_0'(\omega)\tilde r''(\omega)+\tilde G_0''(\omega)\tilde r'(\omega)
-\Gamma'(\omega)\tilde r''(\omega)+\Gamma''(\omega)\tilde r'(\omega),
\end{aligned}
\label{xp_xpp}
\eeq
where the real and imaginary parts of $\tilde r$ are $\tilde r'$ and $\tilde
r''$, respectively.
Using the parity properties of the Fourier transform of a real function
\beq
\begin{aligned}
    \tilde r'(\nu)&=\tilde r'(-\nu),\\
    \tilde r''(\nu) &=-\tilde r''(-\nu),
\end{aligned}
\eeq
and the following statistical averages of white noise in the frequency domain:
\beq
\begin{aligned}
\langle \tilde r'(\nu)\tilde r'(\nu')\rangle &=2\pi
D\left[\delta(\nu-\nu')+\delta(\nu+\nu')\right],\\
\langle \tilde r'(\nu)\tilde r''(\nu')\rangle &=0,\\
\langle \tilde r''(\nu)\tilde r''(\nu')\rangle &=2\pi
D\left[\delta(\nu-\nu')-\delta(\nu+\nu')\right],
\end{aligned}
\eeq
At $\nu=\omega$, we obtain the two dispersions in quadrature and the
correlation to be given by
\beq
\begin{aligned}
\sigma_c^2(\omega)&=\lim_{\Delta\nu\rightarrow 0^+}\int^{\omega+\Delta\nu}_{\omega-\Delta\nu} 
\langle \tilde x'(\omega)\tilde x'(\nu')\rangle\; d\nu'\\
&=
2\pi D\left\{|\tilde G_0(\omega)|^2+|\Gamma(\omega)|^2
+2\Re\{\tilde G_0(\omega)\Gamma(\omega)\}\right\}\\
&=\dfrac{2\pi D\left[\Omega^2+\gamma^2\omega^2+F_p^2/4-\gamma\omega F_p\right]}{\left[\Omega^2+\gamma^2\omega^2-F_p^2/4+\eta(\gamma\omega-F_p/2)\right]^2},
\end{aligned}
\label{sigma2_c}
\eeq
\beq
\begin{aligned}
\sigma_s^2(\omega)&=\lim_{\Delta\nu\rightarrow 0^+}\int^{\omega+\Delta\nu}_{\omega-\Delta\nu} 
\langle \tilde x''(\omega)\tilde x''(\nu')\rangle\; d\nu'\\
&=2\pi D\left\{|\tilde G_0(\omega)|^2+|\Gamma(\omega)|^2
-2\Re\{\tilde G_0(\omega)\Gamma(\omega)\}
\right\}\\
&=\dfrac{2\pi D\left[ \Omega^2+(\gamma\omega+\eta/2)^2+(F_p+\eta)^2/4+(\gamma\omega+\eta/2)(F_p+\eta)\right]}{\left[\Omega^2+\gamma^2\omega^2-F_p^2/4+\eta(\gamma\omega-F_p/2)\right]^2},
\end{aligned}
\label{sigma2_s}
\eeq
and
\beq
\begin{aligned}
\sigma_{cs}(\omega)&=\lim_{\Delta\nu\rightarrow 0^+}\int^{\omega+\Delta\nu}_{\omega-\Delta\nu} 
\langle \tilde x'(\omega)\tilde x''(\nu')\rangle\; d\nu'
=4\pi D \Im\left\{\tilde G_0(\omega)\Gamma(\omega)\right\}\\
&=\dfrac{2\pi D \Omega(F_p+\eta)}{\left[\Omega^2+\gamma^2\omega^2-F_p^2/4+\eta(\gamma\omega-F_p/2)\right]^2},
\label{sigma_cs_om}
\end{aligned}
\eeq
where we used the relations
\beq
\begin{aligned}
\Re\{\tilde G_0(\omega)\Gamma(\omega)\}&=-\dfrac{(\gamma\omega+\eta/2)(F_p+\eta)/2}{\left[\Omega^2+\gamma^2\omega^2-F_p^2/4+\eta(\gamma\omega-F_p/2)\right]^2},\\
\Im\{\tilde G_0(\omega)\Gamma(\omega)\}&=\dfrac{\Omega(F_p+\eta)/2}{\left[\Omega^2+\gamma^2\omega^2-F_p^2/4+\eta(\gamma\omega-F_p/2)\right]^2}.
\end{aligned}
\eeq

In the important case of no detuning ($\Omega=0$), we have
\beq
\begin{aligned}
\sigma_c^2(\omega)&=\dfrac{2\pi D\left[\gamma^2\omega^2+F_p^2/4-\gamma\omega F_p\right]}{\left[\gamma^2\omega^2-F_p^2/4+\eta(\gamma\omega-F_p/2)\right]^2}
=\dfrac{2\pi D}{\left[\gamma\omega+F_p/2+\eta\right]^2},\\
\sigma_s^2(\omega)&=\dfrac{2\pi D\left[(\gamma\omega+\eta/2)^2+(F_p+\eta)^2/4+(\gamma\omega+\eta/2)(F_p+\eta)\right]}{\left[\gamma\omega+F_p/2+\eta\right]^2(\gamma\omega-F_p/2)^2}=\dfrac{2\pi D}{(\gamma\omega-F_p/2)^2},\\
\sigma_{cs}(\omega)&=0.
\label{sigmas_om1}
\end{aligned}
\eeq
We can normalize these fluctuations by dividing them by the value $\sigma_0^2=2\pi D/(\gamma\omega)^2$ when $F_p=\eta=0$.
We find
\beq
\begin{aligned}
\sigma_c^2(\omega)&=\dfrac{\sigma_0^2}{\left(1+r+g\right)^2},\\
\sigma_s^2(\omega)&=\dfrac{\sigma_0^2}{(1-r)^2},\\
\label{sigmas_normalized}
\end{aligned}
\eeq
where $r=F_p/(2\gamma\omega)$ and $g=\eta/(\gamma\omega)$.
This result is equivalent to the one obtained in Eqs.~\eqref{r_min_r_max}.
From the equations \eqref{sigmas_normalized}, we conclude that one can reach
squeezing of $\sigma_c$ as strong as possible by increasing the value of $\eta$
with $F_p=0$ while $\sigma_s$ remains fixed at $\sigma_0$.
One limitation of this method is due to the fact that when increasing $\eta$
the averaging method eventually breaks down.
Furthermore, the approximation that led to equation \eqref{noise_response} also breaks down
since one has to take into account the off-resonance terms with $\tilde x(\pm 3\omega)$ in
Eq.~\eqref{polf_noise_fourier}.
For very high $Q$ resonators, the breakdown of this approximation is is a minor
issue though.
With $r=0$, the maximum gain is 1 and the squeezing in the cosine quadrature
has no limit what leads to an overall cooling of the resonator. 
When $g=0$ instead, the strongest squeezing occurs at $r\rightarrow1$ with the
lower limit being $-6\si{dB}$, what is in agreement with
results of Rugar and Grutter \cite{rugar91} and Cleland
\cite{cleland2005thermomechanical}, but in disagreement with the $-3\si{~dB}$
from Vinante and Falferi \cite{vinante2013feedback} and Poot \etal
\cite{poot2015deep}.
\subsubsection{Noise spectral density}
We can write Eq.~\eqref{polf_noise_fourier} as
\beq
\begin{aligned}
&\left\{1
-\frac{\eta\chi(\nu)}{2\tau}\left[\frac{1-e^{i(\nu+\omega)\tau}
}{\nu+\omega}
-\frac{1-e^{i(\nu-\omega)\tau}}{\nu-\omega}\right]\right\}\tilde x(\nu)\\
&+\frac{\chi(\nu)}{2}\left\{-iF_p
+\frac{\eta}{\tau}
\frac{\left[1-e^{i(\nu-\omega)\tau}\right]}{\nu-\omega}
\right\}\tilde x(\nu-2\omega)\\
&+\frac{\chi(\nu)}{2}\left\{iF_p
-\frac{\eta}{\tau}\frac{1-e^{i(\nu+\omega)\tau}}{\nu+\omega}
\right\}\tilde x(\nu+2\omega)=\chi(\nu)\tilde r(\nu).
\end{aligned}
\eeq
We can rewrite the above equation as
\beq
A(\nu)\tilde x(\nu)+B^*(-\nu)\tilde x(\nu-2\omega)+B(\nu)\tilde x(\nu+2\omega)=\chi(\nu)\tilde r(\nu),
\label{eq:fundamental_noise}
\eeq
where we used the following shorthand expressions
\beq
\begin{aligned}
A(\nu) &=1-\frac{\eta\chi(\nu)}{2\tau}\left[\frac{1-e^{i(\nu+\omega)\tau}
}{\nu+\omega}
-\frac{1-e^{i(\nu-\omega)\tau}}{\nu-\omega}\right],\\
B(\nu) &= \frac{\chi(\nu)}{2}\left[iF_p
-\frac{\eta}{\tau}\frac{1-e^{i(\nu+\omega)\tau}}{\nu+\omega}
\right].
\end{aligned}
\eeq
Near resonance ($\nu\approx\omega\approx\omega_0$), we can find an approximate
closed system of equations
\beq
\begin{aligned}
A(\nu-2\omega)\tilde x(\nu-2\omega)+B(\nu-2\omega)\tilde x(\nu)&=
\chi(\nu-2\omega)\tilde r(\nu-2\omega),\\
A(\nu+2\omega)\tilde x(\nu+2\omega)+B^*(-\nu-2\omega)\tilde x(\nu)&=\chi(\nu+2\omega)\tilde r(\nu+2\omega),
\end{aligned}
\eeq
where we neglected terms with $\tilde x(\nu\pm4\omega)$.
With the help of these equations, we can write $\tilde x(\nu-2\omega)$ and $\tilde x(\nu+2\omega)$ in terms of $\tilde x(\nu)$, $\tilde r(\nu-2\omega)$, and
$\tilde r(\nu+2\omega)$ as
\beq
\begin{aligned}
\tilde x(\nu-2\omega)&=\frac1{A(\nu-2\omega)}\left[-B(\nu-2\omega)\tilde x(\nu)
+\chi(\nu-2\omega)\tilde r(\nu-2\omega)\right],\\
\tilde x(\nu+2\omega)&=\frac1{A(\nu+2\omega)}\left[-B^*(-\nu-2\omega)\tilde x(\nu)+\chi(\nu+2\omega)\tilde r(\nu+2\omega)\right].
\end{aligned}
\eeq
We replace $\tilde x(\nu-2\omega)$ and $\tilde x(\nu+2\omega)$ in
Eq.~\eqref{eq:fundamental_noise} to obtain
\beq
\begin{aligned}
&\left[A(\nu)-\frac{B^*(-\nu)B(\nu-2\omega)}{A(\nu-2\omega)}
-\frac{B(\nu)B^*(-\nu-2\omega)}{A(\nu+2\omega)}\right]\tilde x(\nu)=\\
&\chi(\nu)\tilde r(\nu)-\frac{B^*(-\nu)\chi(\nu-2\omega)}{A(\nu-2\omega)}
\tilde r(\nu-2\omega)
-\frac{B(\nu)\chi(\nu+2\omega)}{A(\nu+2\omega)}\tilde r(\nu+2\omega).
\end{aligned}
\eeq
We can recast the approximate solution above in terms of elastic scattering,
down, and up-convertions of the input noise as
\beq
\tilde x(\nu)=\mg_0(\nu)\tilde r(\nu)+\mg_+(\nu)\tilde r(\nu-2\omega)+\mg_-(\nu)\tilde r(\nu+2\omega),
\label{x_nu}
\eeq
where
\beq
\begin{aligned}
\mg_0(\nu) &= \frac{\chi(\nu)}{A(\nu)-\frac{B^*(-\nu)B(\nu-2\omega)}{A(\nu-2\omega)}
-\frac{B(\nu)B^*(-\nu-2\omega)}{A(\nu+2\omega)}},\\
\mg_+(\nu) &= -\frac{B^*(-\nu)\chi(\nu-2\omega)}{A(\nu-2\omega)\left[A(\nu)-\frac{B^*(-\nu)B(\nu-2\omega)}{A(\nu-2\omega)}
-\frac{B(\nu)B^*(-\nu-2\omega)}{A(\nu+2\omega)}\right]}, \\
\mg_-(\nu) &= -\frac{B(\nu)\chi(\nu+2\omega)}{A(\nu+2\omega)\left[A(\nu)-\frac{B^*(-\nu)B(\nu-2\omega)}{A(\nu-2\omega)}
-\frac{B(\nu)B^*(-\nu-2\omega)}{A(\nu+2\omega)}\right]}.
\end{aligned}
\eeq
The NSD $S_{\tilde x}$ as defined in Ref.~\cite{batista2022gain} is given by
\beq
S_{\tilde x}(\nu)=\lim_{\Delta\nu\rightarrow 0^+}\int^{\nu+\Delta\nu}_{\nu-\Delta\nu} 
\dfrac{\langle \tilde x(-\nu)\tilde x(\nu')\rangle}{2\pi}\; d\nu'.
\label{S_N}
\eeq
When $\nu\neq\omega$, we find
\beq
S_{\tilde x}(\nu)=2D\left[|\mg_0(\nu)|^2+|\mg_+(\nu)|^2+|\mg_-(\nu)|^2\right],
\label{S_Nnu}
\eeq
where we used the fact that $\tilde r(\nu)$ is
a white noise with zero mean and the statistical average
$\langle\tilde r(\nu)\tilde r(\nu')\rangle=4\pi D\delta(\nu+\nu')$.
\section{Results and discussion}
\label{results}
\begin{figure}[!ht]
    \centerline{\includegraphics[{scale=0.8}]{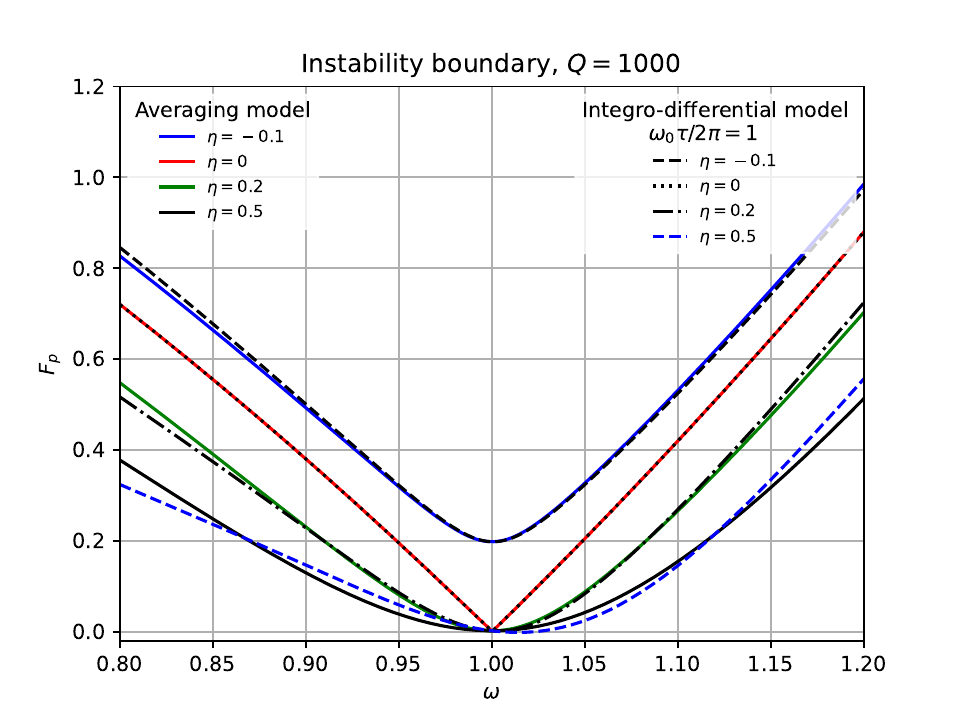}}
\caption{Parametric instability thresholds for several values of the feedback
constant $\eta$. 
The solid lines are obtained from Eq.~\eqref{threshold}, whereas the
dashed or dotted lines are given by $|a(\omega)|=|b(\omega)|$ corresponding to
the zero of the denominator of Eq.~\eqref{AomF_s} for the fixed value of 
the lock-in time constant $\tau$ shown in the figure.
}
\label{fig:threshold}
\end{figure}

In Fig.~\ref{fig:threshold} we plot the threshold lines of instability for
several values of $\eta$ in the first-order averaging approximation, given by
Eq.~\eqref{threshold} (solid lines), and of the integro-differential model,
(discontinuous lines) which correspond to the zero of the denominator of
Eq.~\eqref{AomF_s}. 
The discrepancy between the two models are due to the fact that implicitly
the lock-in time constant is varied in the averaging model so that $\tau$ is
an integer multiple of $2\pi/\omega$, whereas in the integro-differential model
we use a fixed value of $\tau$ $2\pi/\omega_0$.
The latter case is easier to perform experimentally and also more meaningful,
since otherwise one has to vary three parameters ($\omega,\, F_p,$ and $\tau$),
instead of only two ($\omega$ and $F_p$), to trace the threshold line. 
It is noteworthy to mention that once there is feedback, the inversion symmetry
$F_p\rightarrow-F_p$ is broken for the threshold lines. 
Because of this, the threshold line with $\eta=0.5$ crosses $F_p=0$ just above
$\omega=1$.

\begin{figure}[!ht]
    \centerline{\includegraphics[{scale=0.8}]{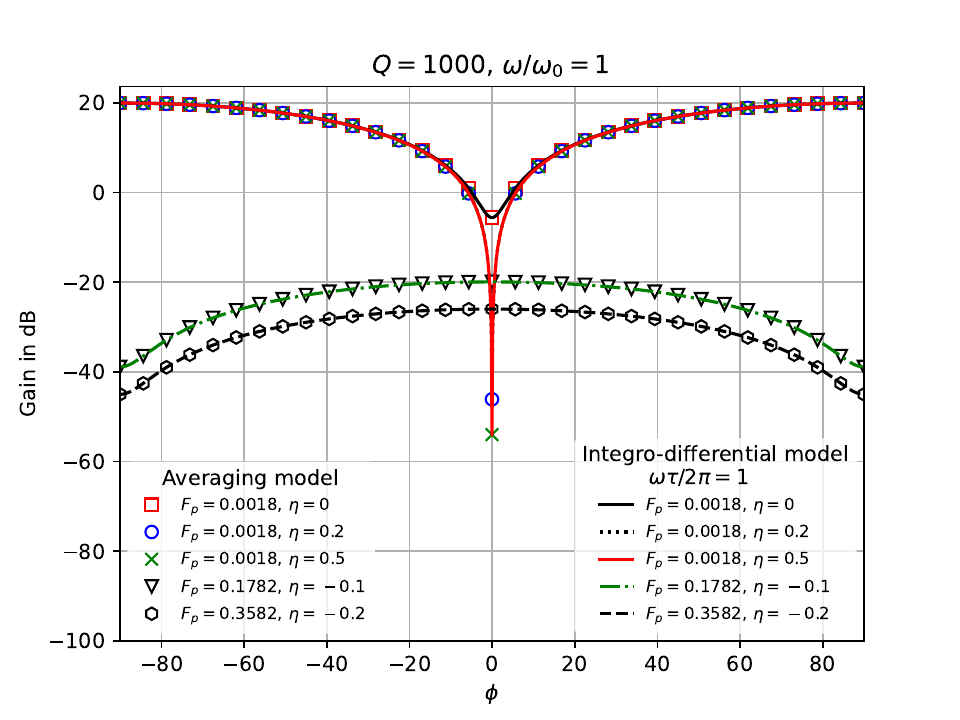}}
\caption{Gain as a function of phase $\phi$ for a parametric amplifier with
lock-in feedback.
We plot the gain given by $20\log_{10}\frac{r_{F_p,\eta}}{r_0}$, where the
amplitude $r_{F_p,\eta}$ is the response of the parametric resonator with
lock-in feedback and $r_0$ is the harmonic resonator response
amplitude. 
The averaging model results are obtained from Eq.~\eqref{deep_squeeze}.
The integro-differential model results are obtained from Eq.~\eqref{AomF_s}.
All results with feedback yield deep deamplification (below -6 dB) at least
in a narrow range around $\phi=0$ for $\eta>0$ and for all phases when $\eta<0$.
}
\label{fig:gain}
\end{figure}

\begin{figure}[!ht]
    \centerline{\includegraphics[{scale=0.8}]{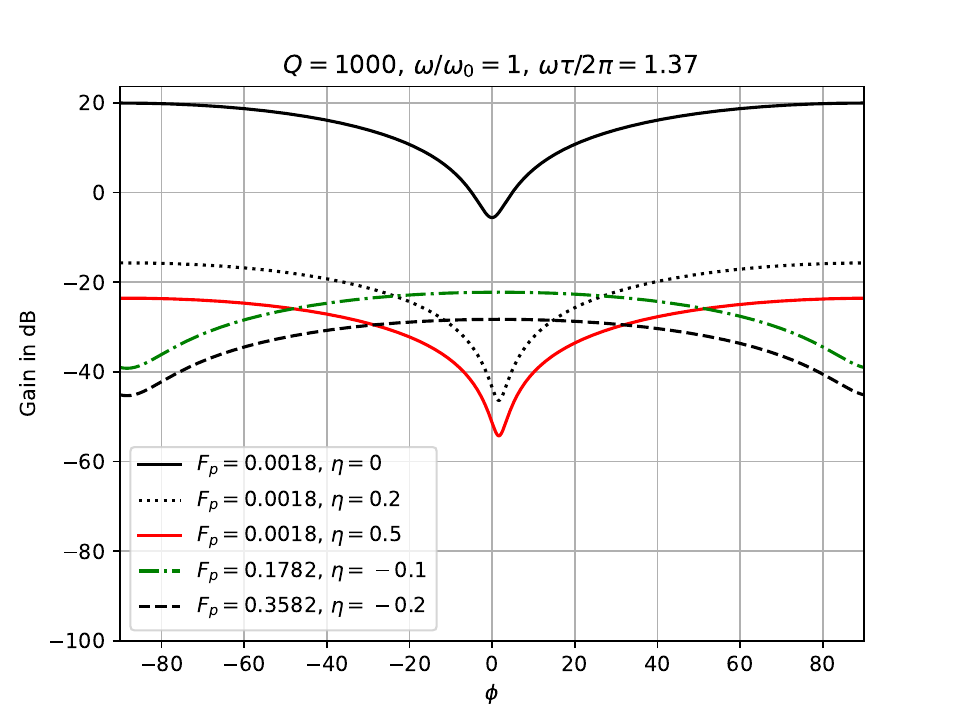}}
\caption{
Gain as a function of phase $\phi$ for a parametric amplifier with
lock-in feedback.
We plot the gain $20\log_{10}\frac{r_{F_p,\eta}}{r_0}$ in decibels, where the
amplitude $r_{F_p,\eta}$ is the response of the parametric resonator with
lock-in feedback obtained from Eq.~\eqref{AomF_s} with $\omega\tau/2\pi=1.37$
and $r_0$ is the harmonic resonator response amplitude. 
We obtain considerably smaller gains than obtained in Fig.~\ref{fig:gain}
with the same parameters except for the value of $\tau$.
For $\eta=0.2$ and $0.5$, one obtains deamplification in all phases.
    }
\label{fig:gain_tau}
\end{figure}

In Fig.~\ref{fig:gain} we plot the degenerate parametric amplification gain in
dB with respect to the simple forced harmonic resonator response as a function
of $\phi$.
As expected, when the lock-in time-constant $\tau$ is a multiple of
$2\pi/\omega$ both models predict the same gain.
In Fig.~\ref{fig:gain_tau} we likewise plot the degenerate parametric
amplification gain in dB with respect to the simple forced harmonic resonator
response but with $\omega\tau/2\pi=1.37$.  
These results cannot be reproduced by the averaged feedback model of Eq.~\eqref{threshold}.
For the same values of $F_p$ and $\eta$ we obtain considerably stronger
cooling, especially for the $\eta=0.2$ and $\eta=0.5$ curves.
All pump amplitudes in both these figures are set at 90\% of the threshold
value.

\begin{figure}[!ht]
    \centerline{\includegraphics[{scale=0.4}]{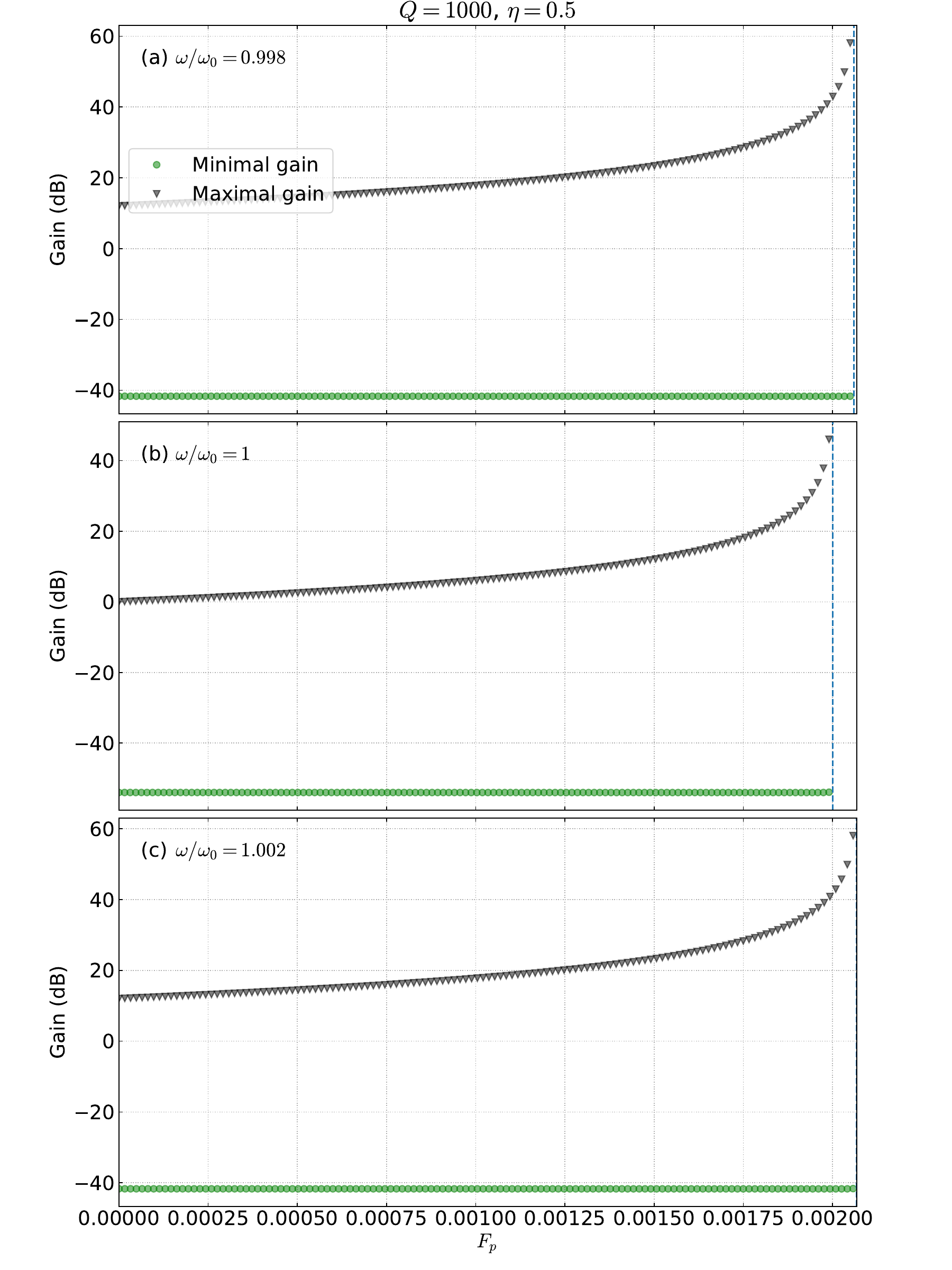}}
\caption{Minimal and maximal gains as a function of pump amplitude.
Here we use the same gain expression as in Fig.~\ref{fig:gain}.}
\label{fig:min_max_gains}
\end{figure}

\begin{figure}[!ht]
    \centerline{\includegraphics[{scale=0.4}]{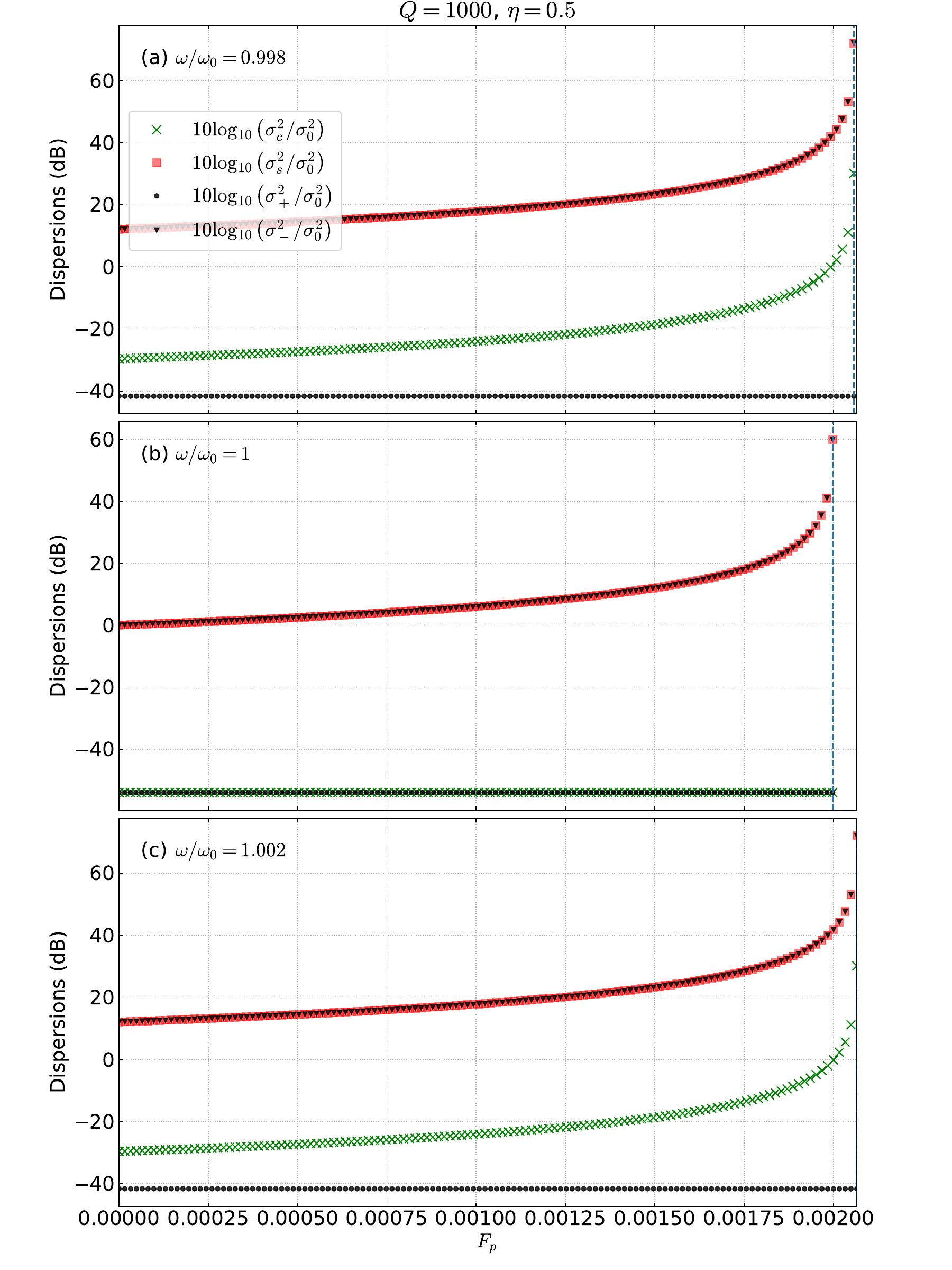}}
\caption{Standard deviations: cosine quadrature ($\sigma_c^2$ from
Eq.~\eqref{sigma2_c}), sine quadratude ($\sigma_s^2$ from
Eq.~\eqref{sigma2_s}), and eigenvalues of the diagonalized covariance matrix
($\sigma_-^2$ and $\sigma_+^2$) in dB scale relative to the equilibrium values
obtained at zero pump and zero feedback.
Each plot ends very close to the instability threshold (vertical dashed lines). 
The common parameters used in all panels are given on top of the figure.
In all panels we see that, unlike the single-degree-of-freedom
parametric resonator without feedback, deep squeezing far below $-6~$dB can be achieved in the parametric resonator with feedback.
The squeezing is practically independent of parametric modulation, but
decreases with detuning.
The diagonalized standard deviations are in agreement with the maximal and minimal gains depicted in
Fig.~\ref{fig:min_max_gains}.
This indicates that our stochastic model is consistent with phase-dependent
parametric amplification.
}
\label{fig:squeezingFB}
\end{figure}

In Fig.~\ref{fig:min_max_gains} we plot the minimal and maximal degenerate
parametric amplification gain obtained from Eq.~\eqref{deep_squeeze} as a
function of pump amplitude with (a) $\omega/\omega_0=0.998$, (b) $\omega/\omega_0=1$, (c) $\omega/\omega_0=1.002$. 
As a comparison we plot in Fig.~\ref{fig:squeezingFB} the dispersions
$\sigma_c^2$ and $\sigma_s^2$ along with the diagonalized dispersions
$\sigma_-^2$ and $\sigma_+^2$ as a function of $F_p$ with $\omega\tau/2\pi=1$.
For details on obtaining $\sigma_-$ and $\sigma_+$ see Ref.~\cite{batista2024deep}.
As expected, the diagonalized dispersions behave very closely with the results
plotted in Fig.~\ref{fig:min_max_gains}.
 
\begin{figure}[!ht]
    \centerline{\includegraphics[{scale=0.6}]{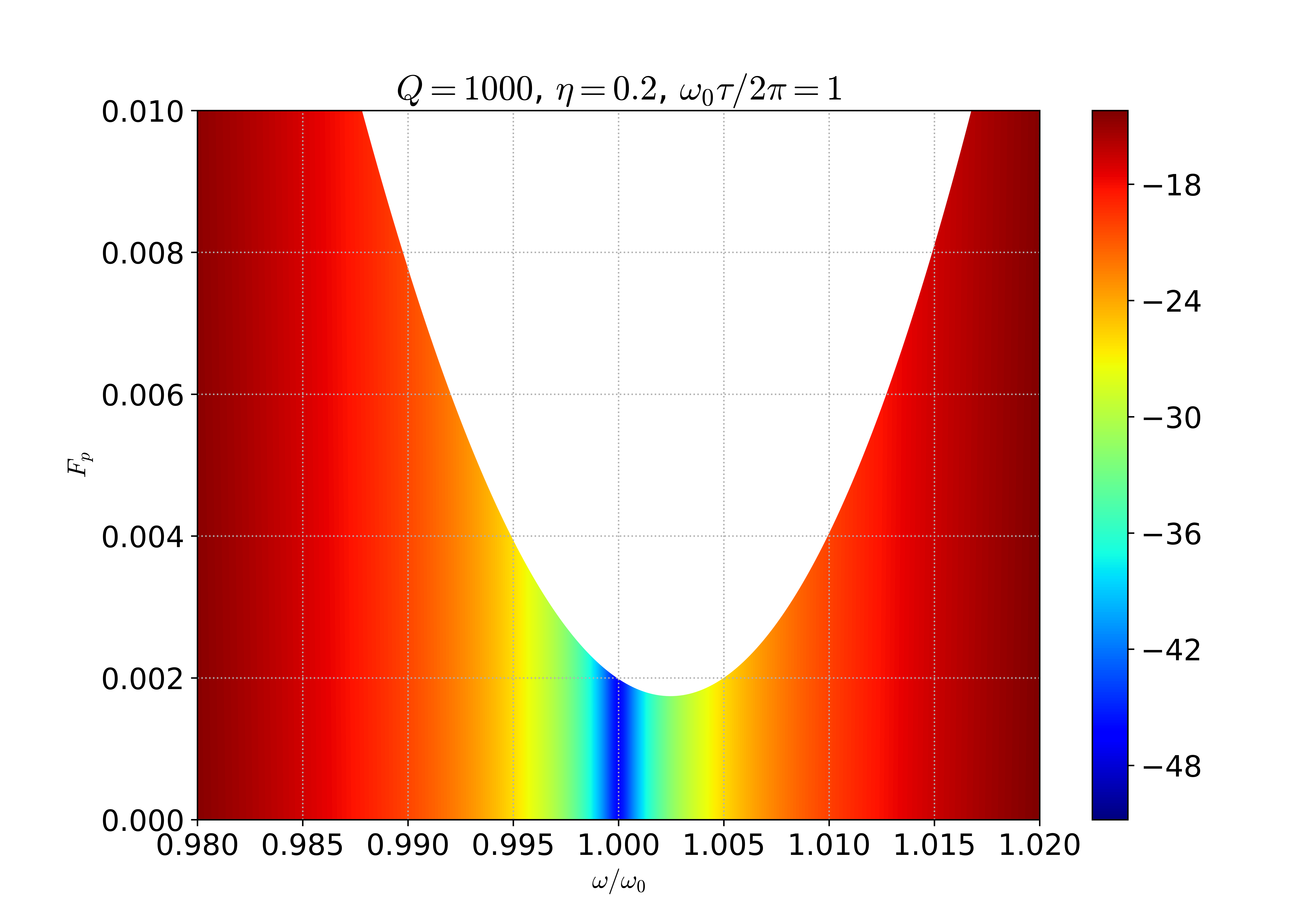}}
\caption{Lock-in feedback minimum dispersion colormap in dB of the
parametrically-driven amplifier. 
At each point in $(\omega, F_p)$ parameter space we plot the smallest
dispersion obtained from the diagonalization of the covariance matrix with
elements given in Eq.~\eqref{sigma2_c}-\eqref{sigma_cs_om}.}
\label{fig:sigma_2m}
\end{figure}

In Fig.~\ref{fig:sigma_2m} we plot a color map of the strongest squeezing
after diagonalization of the two-variable Gaussian probability distribution
that generates the dispersions and correlation given in
Eqs.~\eqref{sigma2_c}-\eqref{sigma_cs_om}.
The strongest squeezing occurs in a narrow band around $\omega=\omega_0$.
The value of $\tau$ is held constant at $\omega_0\tau/2\pi=1$.
As expected, there is little dependence of the squeezing on $F_p$ since $g=500$
and $r=1$ at threshold from Eq.~\eqref{sigmas_normalized}.

\begin{figure}[!ht]
    \centerline{\includegraphics[{scale=0.6}]{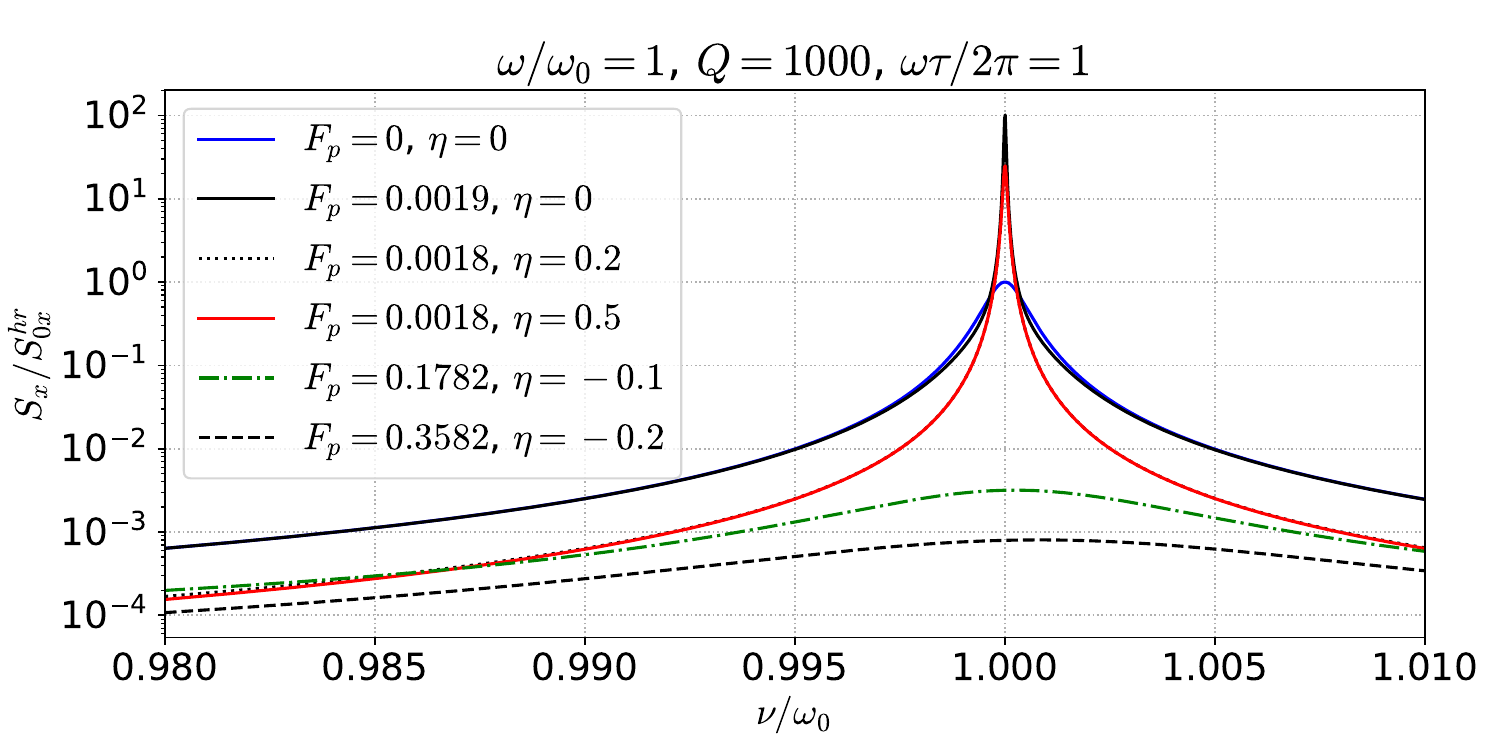}}
\caption{
Normalized noise spectral density lines obtained from Eq.~\eqref{S_Nnu} with
$\tau=2\pi/\omega$ for several values of $F_p$ and $\eta$.
All NSD's are dividided by the peak value of the harmonic resonator NSD.
The blue line is the NSD of the harmonic resonator, the solid black line
corresponds to the NSD of the parametric resonator.
The remaining lines correspond the parametric resonator with feedback.
For $\eta>0$, we see that the feedback does not alter the NSD of the parametric
resonator near resonance in any significant way.
On the other hand, far from resonance, the feedback response overules the
parametric response.
As a comparison, the peak value of the curve with $F_p=0$ and $\eta=0.5$ is
about $1/4$ of the peak value of the NSD of the harmonic resonator. 
There is substantial cooling only when $\eta<0$.
}
\label{fig:S_Nnu}
\end{figure}

\begin{figure}[!ht]
    \centerline{\includegraphics[{scale=0.6}]{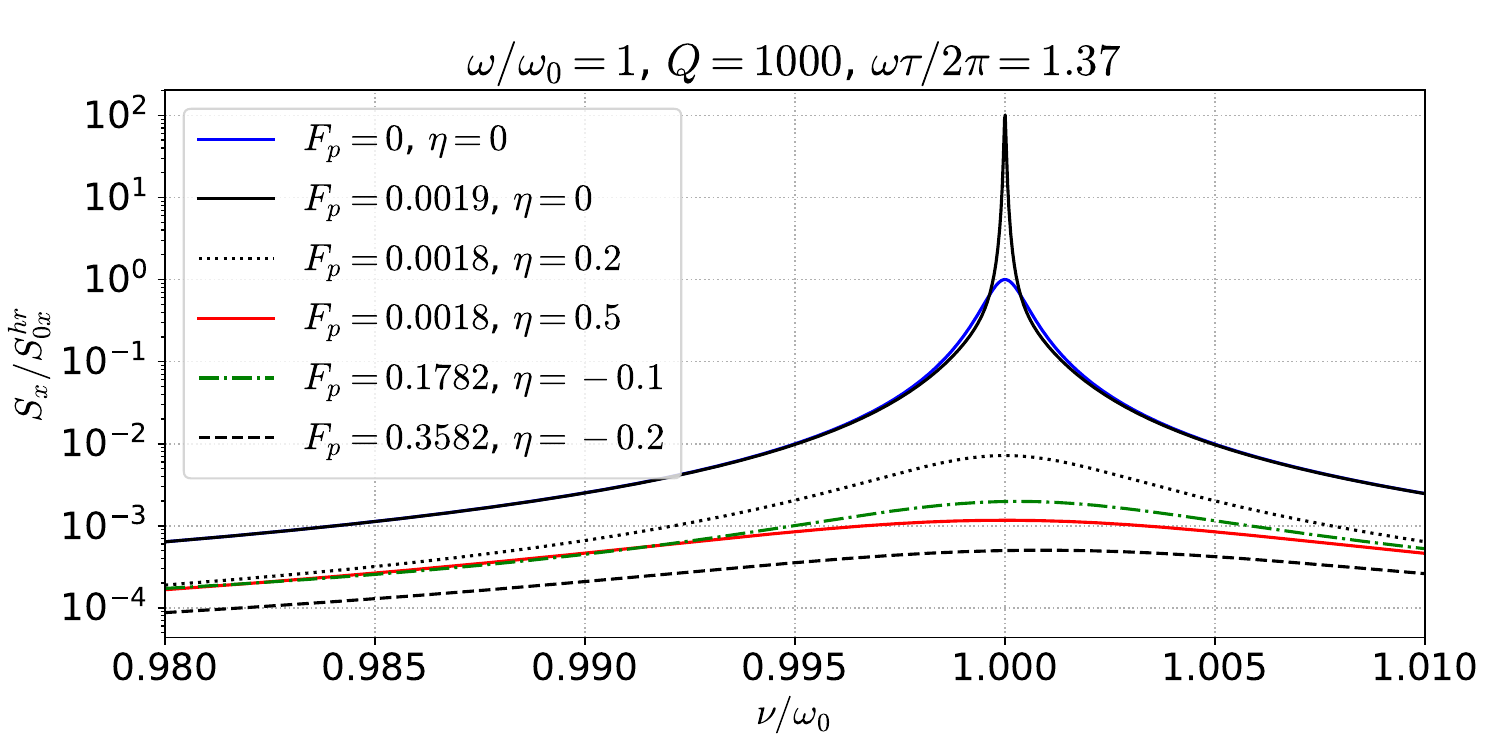}}
\caption{
Normalized noise spectral density curves with $\omega\tau/2\pi=1.37$ for several
values of $F_p$ and $\eta$.
The blue line is the NSD of the harmonic resonator, the solid black line
corresponds to the NSD of the parametric resonator.
We observe that there is cooling when there is feedback.
When $\eta=0.5$, the peak value of this curve is about $1.04\times10^{-3}$ of
the peak value of the NSD of the harmonic resonator. 
Even stronger cooling occurs when $\eta<0$, but at the expense that $F_p$ is
set at a much higher value. 
These represent a far more substantial noise reduction than the ones achieved
by the feedback scheme with $\tau=2\pi/\omega$ plotted in Fig. \ref{fig:S_Nnu}.
}
\label{fig:S_Nnu2}
\end{figure}

In Fig.~\ref{fig:S_Nnu} we plot the noise spectral density given by
Eq.~\eqref{S_Nnu} for the variate $\tilde x(\nu)$ obtained from the Fourier
transform of the integro-differential equation \eqref{polf_noise} with
$\tau=2\pi/\omega$. 
These results are beyond the scope of the model developed by Vinante and
Falferi.
Indeed, to the author's knowledge, this result is new in the scientific
literature.
We believe this measurement of the NSD is an important test for the theory,
even more so than the squeezing estimates, since with it one can estimate the
amount of cooling.
Also, the squeezing measurements are made only at $\nu=\omega$, whereas the
NSD is obtained across the spectrum of the resonator's response.
In Fig.~\ref{fig:S_Nnu2} we find considerably stronger cooling for the same
parameters $F_p$ and $\eta$ as in the previous figure.
We chose the value $\omega\tau/2\pi=1.37$ that resulted in the strongest
cooling.

The model of Vinante and Falferi is based on the application of
the averaging method to a single-degree-of-freedom degenerate parametric
amplifier at resonance ($\omega=\omega_0$).
In their analysis, the authors replaced the single tone external drive by white
noise and, subsequently, Fourier transformed the averaged equations.
To our knowledge, there are several mathematical problems with this approach.
TThe averaging method is usually applied to transform non-autonomous dynamical
systems into slowly varying autonomous dynamical systems.
For instance, if there is a zero-mean time-periodic parametric pump with period $\pi/\omega$
with $\omega\approx\omega_0$ in the original driven resonator, then only the
Fourier component of the parametric pump at $2\omega$ remain in
the first-order averaged dynamical system, all higher harmonics are
eliminated.
If the resonator is also additively driven by a time periodic signal with
period $2\pi/\omega$, then after averaging only the first Fourier component at
$\omega$ will play a role, with all higher harmonics filtered out.
Hence, the application of the averaging method works as a band-pass filter at 
$\omega\approx\omega_0$.
So it seems there was an inconsistency in the analysis developed by Vinante and
Falferi when applying the averaging method to a dynamical system driven by
white noise. 
In their model, the averaged noise is still a white noise, with Fourier
components with the same amplitude accross the spectrum.
In reality, the averaged noise should be a correlated noise.
The Fourier analysis of an averaged dynamical system also seems contradictory,
since high Fourier components will correspond to terms varying exceedingly fast in time.
Finally, the averaged slow quadrature variables ($X$ and $Y$ in their notation)
are related to the Fourier peak at $\omega_0$ of $\tilde x$, that is $|\tilde
x(\omega_0)|=\sqrt{X^2+Y^2}$), hence the Fourier transform of $X(t)$ and $Y(t)$
does not seem mathematically meaningful.
Here we avoided these issues by directly Fourier transforming Eq.~\eqref{polf_noise} into the frequency domain, obtaining Eq.~\eqref{polf_noise_fourier}, which
is exact.
We proceeded by neglecting off-resonance terms such as $\tilde
x(\pm3\omega)$ in Eq.~\eqref{xom_fb} to obtain squeezing or $\tilde x(\nu\pm4\omega)$ in Eq.~\eqref{x_nu} in order to obtain an approximate expression for the NSD.

\FloatBarrier

\FloatBarrier
\FloatBarrier
\section{Conclusion}
\label{conclusion}
Here we developed a linear feedback scheme that is able to achieve squeezing
and cooling of fluctuations in a resonator.
The feedback scheme we proposed is described by an integro-differential
equation, which in the most general form consists of an integral of the
response of the resonator multiplied by a cosine and integrated over a given
time interval.
We believe our feedback scheme is more physical than the one proposed by
Vinante and Falferi since it emulates the behavior of a lock-in amplifier.
We can also easily account for a linear combination of $u$ and $v$ (the slowly
varying averaged variables) in the feedback by changing the integration time
$\tau$ of the lock-in amplifier.
By doing so, we were able to achieve deeper squeezing and also stronger cooling
for the same feedback constant than in Vinante and Falferi scheme.

Initially, we analysed the response of the resonator with feedback to an added
ac signal without noise.
When this feedback is applied to a harmonic resonator, it deamplifies its
response with respect to the pure harmonic resonator response.
When applied to a parametric resonator, one can obtain a strong dependence on
phase, in some cases with amplification in one quadrature and deep
deamplification in the other quadrature.
Depending on the value of the feedback constant, one can also obtain
deamplification in both quadratures, what leads to cooling.
We verified that the phase-dependent gain obtained in Eq. \eqref{polf_gain} and
Eq.~\eqref{polf_gain_integral} were the same when the lock-in time constant
$\tau$ is a multiple of half the pump period.

We also investigated the response of the resonator (harmonic or parametric)
with feedback to added white noise.
Our model avoids the application of the averaging method in the presence of
added noise.
Instead, we analysed the stationary fluctuations of the resonator in the
frequency domain.
We observed that we can achieve deep squeezing far below the $-6$dB limit of
pure parametric squeezing set by Rugar and Grutter's model.
With our model, we were able to compute not only the squeezing at half the pump
frequency, but also the NSD across a frequency band around resonance.
We only added noise to the integro-differential model to avoid the issues of
averaged dynamical systems with added noise.
Instead of applying the averaging method, we Fourier analysed the original equations of motion
\eqref{polf_gain_integral}.

In addition to squeezing, one can also obtain very strong cooling as evidenced
by the reduction of the NSD by a factor of up to $10^4$.
The calculation of the NSD allows for further experimental tests of our model.
It would be very interesting to verify the amount of parameter space that is
available experimentally.


%
\end{document}